\documentclass{article}
\usepackage{amsmath,amsthm,stmaryrd,graphicx}
\usepackage[all]{xy}

\title{%
  Tuplix Calculus Specifications
  of Financial Transfer Networks}
\author{%
	Jan A.\ Bergstra$^{1}$\thanks{
	  Partially supported by the
	  Dutch NWO Jaquard Project
	  Symbiosis, project number 638.003.611.}
	   \and
	Sanne Nolst Trenit\'e$^{2}$\and
	Mark B.\ van der Zwaag$^{1}$
\\
  {\small
	  $^{1}$Section Software Engineering,
	  Informatics Institute,
	  University of Amsterdam}\\
  {\small
	  $^{2}$Faculty of Science,
	  University of Amsterdam}\\
  {\small Email: \{janb,sanne,mbz\}@science.uva.nl}
}
\date{}

\theoremstyle{definition}
\newtheorem{definition}{Definition}
\newtheorem{example}{Example}

\newcommand{\axname}[1]{\ensuremath{\textup{#1}}}
\newcommand{\defeq}{\ensuremath{\stackrel{\text{\tiny{def}}}{=}}}

\newcommand{\DE}{\ensuremath{\axname{\textsc{De}}}}
\newcommand{\Attr}{\ensuremath{\mathit{A}}}
\newcommand{\et}{\ensuremath{\varepsilon}} 
\newcommand{\nt}{\ensuremath{\delta}}
\newcommand{\conjc}{\ensuremath{\varobar}}
\newcommand{\Conjc}{\ensuremath{\mbox{\scalebox{1.5}{$\varobar$}}}}
\newcommand{\ztest}[1]{\ensuremath{\gamma(#1)}}
\newcommand{\nztest}[1]{\ensuremath{\widetilde{\gamma}(#1)}}
\newcommand{\FV}{\ensuremath{\mathit{FV}}}
\newcommand{\V}{\ensuremath{\mathit{Var}}}
\newcommand{\gsum}{\ensuremath{\textstyle{\sum}}}
\newcommand{\Clr}{\ensuremath{\et}}
\newcommand{\Enc}{\ensuremath{\partial}}
\newcommand{\Select}{\ensuremath{\mathit{Select}}}
\newcommand{\DM}{\ensuremath{\mathcal{D}} }

\newcommand{\reward}{\ensuremath{\mathit{rew}}}
\newcommand{\pw}{\ensuremath{\mathit{pw}}}
\newcommand{\inc}{\ensuremath{\mathit{inc}}}

\newcommand{\FCO}{\ensuremath{\mathalpha{K}}}
\newcommand{\ATT}{\ensuremath{\mathit{Attr}}}
\newcommand{\UNITS}{\ensuremath{\mathit{Unit}}}
\newcommand{\IN}{\ensuremath{\textit{in}}}
\newcommand{\OUT}{\ensuremath{\textit{out}}}
\newcommand{\scopy}{\ensuremath{\zeta}}

\begin{document}

\maketitle

\begin{abstract}                                   
We study the application of Tuplix Calculus
in modular financial budget design.
We formalize organizational structure
using financial transfer networks.
We consider the notion of flux of money
over a network, and a way to
enforce the matching of influx and outflux for parts
of a network.
We exploit so-called signed attribute notation
to make internal streams visible through encapsulations.
Finally, we propose a Tuplix Calculus construct
for the definition of data functions.
\end{abstract}

\tableofcontents

\section{Introduction}
In~\cite{TFB} we
described the application of Tuplix Calculus
(TC, see~\cite{TC})
in the formalization of financial budgets.
Here, we explore this application further
starting with the
definition of financial transfer networks.
We consider the notion of flux of money
over a network, and define a
{flux constraint operator} that
enforces matching influx and outflux for units.
We exploit so-called signed attribute notation
to make internal streams visible through encapsulations.
Finally, we propose a Tuplix Calculus construct
for the definition of data functions.
We assume familiarity with Tuplix Calculus;
its syntax and axioms are
collected in Appendix~\ref{sec:TC}.

\section{Financial Transfer Networks}
Implicit starting point
in the modular budget design in~\cite{TFB}
is the assumption of an underlying 
(organizational) structure:
tuplix expressions specify budgets
for certain parties, and
by composition we obtain
budgets for larger parts (of an organization).
Of importance is also the identification
of attributes, that are used
in the specification of payments between parts,
or between parts and external parties.
 
\begin{example}
As a simple example, consider an 
organization consisting of parts $P$ and $Q$,
and assume that attribute $a$
is used to specify payments between these parts.
Using the names $P$ and $Q$ also 
as tuplix meta-variables, we define
\[ P = a(10),\quad Q = a(-10).
\]
So, $P$ will pay amount $10$, while
$Q$ intends to receive amount $10$.
When we compose $P$ and $Q$, expressed as
$\Enc_{\{a\}}(P\conjc Q)$,
these entries synchronize successfully.
\end{example}

We find it worthwhile
to introduce a mathematical format for
organizational structures.
We define a \emph{financial transfer network} (FTN)
as a set of \emph{units} with in-going and outgoing
channels:
a channel is a directed link between units,
or between a unit and an external party,
that is labeled with an attribute. 
Labels of in-going channels of a unit are used
in the specification of payments to the unit,
and the labels of outgoing channels are used to
specify payments made by the unit.
We require that any channel is
in-going for at most one unit and outgoing for
at most one unit.

\begin{definition}
An FTN consists of:
\begin{enumerate}
\item a set \ATT\ of attributes;
\item a set \UNITS\ of units;
\item a function $\IN: \UNITS\to 2^{\ATT}$;
\item a function $\OUT: \UNITS\to 2^{\ATT}$;
\end{enumerate}
such that for all distinct $g,h\in\UNITS$,
 $\IN(g)\cap\IN(h)=\emptyset$
 and
 $\OUT(g)\cap\OUT(h)=\emptyset$.

An attribute $a$ is \emph{internal\/}
if there are units $g,h\in\UNITS$ with
$a\in\IN(g)\cap\OUT(h)$.
An attribute is \emph{external\/} if it is not internal.
\end{definition}

An FTN can be depicted in a graph-like manner,
with units as nodes, and
arrows (called channels) labeled with
attributes between units,
or between a unit and an external party.
Because an attribute of an FTN can be the label
of at most one channel, we shall also
speak of the channel $a$, rather than
the channel labeled with attribute $a$.
A channel is internal if its label is internal;
this is the case if it connects units of the network,
see the following example.

\begin{example}
Consider the FTN with 
$\ATT=\{a,b,c\}$, $\UNITS=\{g,h\}$, and
\[
  \IN(g) = \{a\},\quad
  \OUT(g) = \IN(h) = \{b\},\quad
  \IN(h) = \{c\}.
\]
This network is depicted as
\[
  {}\xrightarrow{a}g\xrightarrow{b}
   h\xrightarrow{c}{}
\]
The channels $a,c$ are external, $b$ is internal.
\end{example}

Given an FTN, a \emph{specification} 
of a unit $g$ is a
tuplix expression $P_g$ that uses 
only the elements of
$\IN(g)\cup\OUT(g)$ as attributes.

\begin{example}\label{ex:tfb}
This example is a shortened, simplified
version of the example presented
in~\cite{TFB}.
We have added the presentation of
the organizational structure as an FTN.

We consider an FTN as depicted
in the following picture:
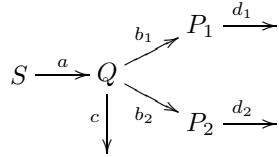
\[
\xymatrix@R=4pt@C=20pt{
 && P_1\ar[r]^{d_1} &\\
 S \ar[r]^{a}&
   Q\ar[dd]_{c}
   \ar[ur]^{b_1}\ar[dr]_{b_2} &&\\
 && P_2\ar[r]^{d_2} &\\
 &
}
\]

The units and their specifications
(for a given period of
 time, e.g., the calendar year 2008):

\begin{itemize}
\item 
$S$ is a \emph{financial source} that
rewards production:
for each product that is produced,
a constant reward $\reward$ is allocated to
unit $Q$.
For production unit $P_i$ (see below)
the data variable $n_i$
stands for the number of products
produced by $P_i$ during the period that is covered.

Specification:
\[ S \defeq a(\reward\cdot(n_1+n_2)).
\]

\item 
The \emph{control\/} unit $Q$ will
dispatch the rewards
to the production units after deduction
of a fixed fraction $k$ (a value between 0 and 1)
that is paid via $c$
to an external service center.
It further distributes the
remainder of the rewards equally among the
production units:
\begin{align*}
 Q\defeq{}&
   \gsum_{x} ( a(-x)\conjc{}\\
& \qquad
    c(k\cdot x)\conjc{}\\
& \qquad
    (1-k)\cdot (b_1(x/2)\conjc b_2(x/2))).
\end{align*}

\item The \emph{production} units
$P_i$, for $i=1,2$,
receive money from $Q$
via $b_i$ and pay for their expenses via $d_i$
(in this simplified example,
these units act as serial buffers only, that
is, they simply pass on what they receive):
\[
 P_i \defeq
   \gsum_{x} ( b_i(-x)\conjc d_i(x) ).
\]
\end{itemize}
A combined budget $B$
is specified by the
encapsulated composition of
these specifications:
\[ B\defeq 
   \Enc_{\{a,b_1,b_2\}}
   (S\conjc Q\conjc P_1\conjc P_2).
\]
The encapsulation
enforces synchronization on the internal channels
and then hides these internal streams
(in Section~\ref{sec:flux} we
elaborate on the notion of streams).

We find (see Appendix~\ref{sec:der} for the derivation):
\begin{align*}
B ={}&
  \gsum_x ( 
   \ztest{x = 
	 \reward\cdot(n_1+n_2)}\conjc{}\\
&\qquad
  c(k\cdot x)\conjc{}\\
&\qquad
  (1-k)\cdot (d_1(x/2)\conjc d_2(x/2))).
\end{align*}

Alternatively, we may redefine $Q$
so that it pays the production units proportionally
to their contribution to the total production:
\begin{align*}
 Q\defeq{}&
   \gsum_{x} ( a(-x)\conjc{}\\
& \qquad
    c(k\cdot x)\conjc{}\\
& \qquad
    (1-k) \cdot x\cdot (b_1(n_1/(n_1+n_2))\conjc
    b_2(n_2/(n_1+n_2)))).
\end{align*}
Then we find, for the combined budget:
\[
B =
 c(k\cdot \reward\cdot (n_1+n_2))\conjc
  (1-k)\cdot (d_1(\reward\cdot n_1)\conjc
    d_2(\reward\cdot n_2))
\]
with a similar derivation.
\end{example}

\section{Flux over a Network}\label{sec:flux}
Unit specifications of an FTN can be thought of
as determining an unrealized flux over the internal
channels of a network.
Take for instance the channel
\[ g\xrightarrow{a}h.
\]
We speak of a \emph{stream} over $a$,
when the total amounts
specified for $a$ by $g$ and by $h$ match
(that is, add up to zero).
We then also say that $g$ has outflux
over $a$ and $h$ has influx over $a$.
When there is no match, there is no flux; the
flux is realized when we 
compose unit specifications,
and encapsulation over the internal attributes
is successful.

A very simple example: consider
\[ g\xrightarrow{a}h
\]
with specifications $P_g=a(t)$ and
$P_h=a(-s)$.
We say that $g$ has outflux of size $t$ along $a$,
and that $h$ has influx of size $s$ along $a$.
If the outflux of $g$ along $a$
matches the influx of $h$ along $a$,
that is, if $t$ equals $s$,
then there is a stream of this size
from $g$ to $h$.
This matching corresponds
to the success of
encapsulation of the composed 
unit specifications:
we find
\[ \Enc_{\{a\}}(P_g\conjc P_h) =\ztest{t=s}.
\]
This encapsulation reduces to an equality test;
unsuccessful encapsulation
yields the null tuplix \nt.
Note that encapsulation
hides the internal transactions;
in Section~\ref{sec:focus}
we look at a way to
make successful internal transactions
(i.e., \emph{flux}) of units visible.

Flux dynamics comes into
play with generalized alternative composition
(summation) over amounts.
For example, redefine
$P_h$ so that it will receive \emph{any} amount,
and send this along:
\[ P_g=a(t),\quad P_h =\gsum_x a(-x)\conjc b(x),
\]
then we find that successful encapsulation
determines the outflux of $h$:
\[ \Enc_{\{a\}}(P_g\conjc P_h) = b(t).
\]

Working with this perspective we find it natural
to be able to require for
certain units that `what goes in also comes out.'
For example,
specify that $h$ will receive any amount
along $a$ and will transfer any amount along $b$:
\[ P_g=a(t),\quad P_h =\gsum_x a(-x)\conjc \gsum_y b(y).
\]
Encapsulation over $a$
will enforce the transfer of amount $t$ along $a$, 
and an additional requirement that the
total flux of $h$ equals zero would
turn $h$ into a serial buffer that
forwards amount $t$ along $b$.

We define a unary
\emph{flux constraint operator}
that does exactly this:
it adds to its argument the constraint
that its total flux equals zero.
This operator (written $\FCO$, after Kirchhoff)
is defined as follows:
\begin{align}
\FCO(X) &= \FCO_0(X)\\
\FCO_t(\nt) &= \nt\\
\FCO_t(\et) &= \ztest{t}\\
\FCO_t(\ztest{x}\conjc X) &= \ztest{x}\conjc\FCO_{t}(X)\\
\FCO_t(a(x)\conjc X) &= a(x)\conjc\FCO_{t+x}(X)\\
\FCO_t(X+Y)&= \FCO_t(X)+\FCO_t(Y)\\
\FCO_t(\gsum_x P) &= \gsum_x(\FCO_t(P)) &x\not\in\FV(t)
\end{align}

\begin{example}\label{ex:resbuff}
We define periodic specifications for a
unit $Q$ and a reserve $R$.
The unit $Q$ receives income from and has expenditures
to external parties. Every period it withdraws a fixed
amount from $R$, and it reserves a fixed percentage
of its income to the reserves of the next period.
Any reserves that are not withdrawn are transferred
to the next period.
The flux constraint operator is
used to enforce this transfer of reserves.
It is also applied to $Q$ so that
it will spend any income that is not reserved.

We make this more precise.
We define $Q_n$ and $R_n$ for the unit $Q$ and 
the reserve $R$ in period $n$.
The following attributes are used:
\begin{itemize}
\item $a_{n+1}$ for the transfer from $R_{n}$ to $R_{n+1}$
\item $b_{n+1}$ for the reservation from $Q_{n}$ to $R_{n+1}$
\item $c_n$ for the withdrawal from $R_{n}$ by $Q_n$
\item $d_n$ for the external income of $Q_n$
\item $e_n$ for the external expenditures of $Q_n$
\end{itemize}

The network is depicted in Figure~\ref{fig:resbuff}.

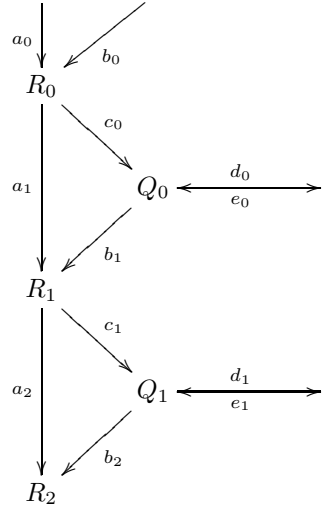
\begin{figure}
\[
\xymatrix{%@R=8pt@C=30pt{
 \ar[d]_{a_0}&\ar[dl]^{b_0}\\
 R_0\ar[dr]^{c_0}\ar[dd]_{a_{1}}
 &&\\
 &Q_0\ar[dl]^{b_{1}}
   \ar@{<->}[rr]_{e_0}^{d_0}&&\\
 R_{1}\ar[dr]^{c_1}\ar[dd]_{a_{2}}
 &&\\
 &Q_1\ar[dl]^{b_{2}}
   \ar@{<->}[rr]_{e_1}^{d_1}&&\\
 R_{2}
 }
\]
\caption{Reserve buffers example}\label{fig:resbuff}
\end{figure}

Define
\begin{align*}
R_n &= \FCO(\gsum_{u,v,w,x} 
  a_n(-u)\conjc 
  b_n(-v)\conjc
  c_n(w)\conjc
  a_{n+1}(x))
\end{align*}
which can be rewritten to
\[
  R_n=
  \gsum_{u,v,w,x} 
  \ztest{u+v=w+x}\conjc
  a_n(-u)\conjc 
  b_n(-v)\conjc
  c_n(w)\conjc
  a_{n+1}(x).
\]

In the specification of $Q_n$ we use
the free data variables
$\pw$ (periodic withdrawal),
$\inc_n$ (income in period $n$),
and
$k$ (reserve fraction, a value between 0 and 1). 
Define
\begin{align*}
Q_n &= \FCO(\gsum_{u} 
  c_n(-\pw)\conjc 
  d_n(-\inc_n)\conjc
  b_{n+1}(k\cdot \inc_n)\conjc
  e_n(u))\\
 &=\gsum_{u}
  \ztest{u=\pw+(1-k)\cdot \inc_n}\conjc{}\\
 &\qquad\qquad  c_n(-\pw)\conjc 
  d_n(-\inc_n)\conjc
  b_{n+1}(k\cdot \inc_n)\conjc
  e_n(u) \\
&= c_n(-\pw)\conjc 
  d_n(-\inc_n)\conjc
  b_{n+1}(k\cdot \inc_n)\conjc
  e_n(\pw+(1-k)\cdot \inc_n) 
\end{align*}

Define
\[ P_n = \Enc_{H_n}(Q_0\conjc\cdots\conjc Q_n\conjc
	R_0\conjc\cdots\conjc R_{n+1})
\] 
where
\[ H_n=\{ a_{i+1},b_{i+1},c_i~|~0\leq i\leq n\}.
\]

For $P_0$ and $P_1$ we find
(see derivations in Section~\ref{sec:der}):
\begin{align*}
P_0 &=
 \FCO(\gsum_{u,v,w,x}{}\\
&\qquad
  a_0(-u)\conjc 
  b_0(-v)\conjc{}\\
&\qquad
  d_0(-\inc_0)\conjc
  e_0(\pw+(1-k)\cdot \inc_0)) \conjc{}\\
&\qquad
  c_1(w)\conjc a_2(x)),
\end{align*}
\begin{align*}
P_1={}&\FCO(\gsum_{u,v,w,x}\\
&\quad
  a_0(-u)\conjc 
  b_0(-v)\conjc{}\\
&\quad  d_0(-\inc_0) \conjc
  e_0(\pw+(1-k)\cdot \inc_0)\conjc{}\\
&\quad
  d_{1}(-\inc_{1})\conjc
  e_{1}(\pw+(1-k)\cdot \inc_{1}) \\
&\quad
  c_{2}(w)\conjc
  a_{3}(x)),
\end{align*}
and this generalizes to
\begin{align*}
P_n={}&\FCO(\gsum_{u,v,w,x}\\
&\quad
  a_0(-u)\conjc 
  b_0(-v)\conjc{}\\
&\quad
  c_{n+1}(w)\conjc
  a_{n+2}(x))\conjc{}\\
&\quad
  \Conjc_{i=0,\ldots,n}\
  d_i(-\inc_i) \conjc
  e_i(\pw+(1-k)\cdot \inc_i)).
\end{align*}
\end{example}

\section{Visualizing Internal Streams}
\label{sec:focus}
In an FTN with unit specifications
we speak of an internal stream over a channel,
if encapsulation over that channel
is successful (does not yield the null tuplix \nt).
In an encapsulation
\[ P = \Enc_H(P_0\conjc\cdots\conjc P_k)
\]
of unit specifications $P_i$, all information 
on internal streams is lost, that is, due
to the encapsulation
no entries with attributes from $H$ occur in $P$.
Still, it may be useful to see the internal streams
of a unit under influence of composition and
encapsulation.  
We shall exploit \emph{signed attribute notation}
to retain focus on encapsulated specifications:
we add copies of internal entries that 
will remain visible after encapsulation.

\subsection*{Signed Attribute Notation}
So far we have used \emph{flat attribute notation}
for entries:
for a unit $g$, if
$a\in\IN(g)$, then an entry $a(t)$ 
is interpreted as influx of amount $-t$ to $g$, and if
$a\in\OUT(g)$, then $a(t)$ 
is interpreted as outflux of amount $t$ from $g$.
The notation is neutral in this respect
(and this is the basis for 
the definition of encapsulation).

An alternative is 
\emph{signed attribute notation}:
for attribute $a$, assume
fresh attributes ${-}a,{+}a$,
and write ${-}a(t)$ for 
influx of amount $t$, and
${+}a(t)$ for outflux of amount $t$.
We have not defined encapsulation
for this notation.

Clearly, tuplix expressions
in signed attribute notation can be transformed
to flat attribute notation
by replacing entries ${+}a(t)$ by $a(t)$,
and ${-}a(t)$ by $a(-t)$.
Vice versa, for a given unit $g$,
transform $a(t)$ to ${-}a(-t)$ if $a\in\IN(g)$, 
and to ${+}a(t)$ if $a\in\OUT(g)$.

\subsection*{Combined Flat and Signed Attribute Notation}
For a unit $g$ and a set of (internal)
attributes $H$, the mapping $\scopy_{g,H}$
will add a
signed copy of internal entries of $g$
in a specification using flat attribute notation.
\begin{align}
\scopy_{g,H}(\nt) &=\nt\\
\scopy_{g,H}(\et) &=\et\\
\scopy_{g,H}(\ztest{x}) &=\ztest{x}\\\displaybreak[0]
\scopy_{g,H}(a(x)) &= 
  \begin{cases}
  {+}a(x)\conjc a(x) &\text{if }a\in\OUT(g)\cap H\\
  {-}a(-x)\conjc a(x) &\text{if }a\in\IN(g)\cap H\\
  a(x)&\text{otherwise}
  \end{cases}\\
\scopy_{g,H}(X\conjc Y) &=\scopy_{g,H}(X)\conjc\scopy_{g,H}(Y)\\
\scopy_{g,H}(X+Y) &=\scopy_{g,H}(X)+\scopy_{g,H}(Y)\\
\scopy_{g,H}(\gsum_xX) &=\gsum_x\scopy_{g,H}(X)
\end{align}
The resulting specification combines
flat and signed attribute notation.

\subsection*{Encapsulation}
Assume we have units $g_0,\ldots, g_k$ with 
corresponding
specifications $P_0,\ldots, P_k$,
and we want to see
what composition and encapsulation
with $P_1,\ldots, P_k$ do to $P_0$.
Let $H$ be the set of attributes that are
internal to $g_0,\ldots,g_k$.
The encapsulation
\[ P = \Enc_H(\scopy_{g_0,H}(P_0)\conjc 
      P_1\conjc\cdots\conjc P_k),
\]
will, if successful, contain
signed copies of the internal transactions of
$g_0$. 
We can now focus on $g_0$ by letting
\[ J = \{ a, {+}a, {-}a~|~a\in\IN(g_0)\cup\OUT(g_0)\},
\]
and selecting (see definition on page~\pageref{def:Select})
on the attributes in this set:
\[ \Select_J(P)
\]
shows all the transactions of $g_0$ under
influence of the encapsulation.

Of course, we can also
make all internal streams of the composition
visible:
\[ \Enc_H(
    \scopy_{g_0,H}(P_0)\conjc 
    \scopy_{g_1,H}(P_1)\conjc\cdots\conjc
    \scopy_{g_k,H}(P_k)).
\]

\begin{example}
Consider the following network:
\[ {}\xrightarrow{a}g\xrightarrow{b}
    h\xrightarrow{c}{}
\]
Take unit specifications
\begin{align*}
P_g &= a(-1)\conjc b(1),\\
P_h &= b(-1)\conjc c(1),
\end{align*}
and observe that
\[ \Enc_{\{b\}}(P_g\conjc P_h) = a(-1)\conjc c(1).
\]
The encapsulation enforces
synchronization on $b$, and leaves
no trace of this synchronization.

Now consider
\[ P = \Enc_{\{b\}}(\scopy_{g,\{b\}}(P_g)\conjc P_h)=
	a(-1)\conjc {+}b(1)\conjc c(1)
\]
where the signed copy 
of the internal outflux of $g$ on $b$
remains visible.
Finally,
let
\[
  J = \{ a, {+}a, {-}a~|~a\in\IN(g)\cup\OUT(g)\},
\]
and find
\[ \Select_J(P)
   =
   a(-1)\conjc {+}b(1).
\]
\end{example}

\section{Function Definition and Binding}
We extend Tuplix Calculus with a construct to define
data functions, 
and with summation over functions.
We only sketch how this extension can be achieved;
a fully worked-out technical account is future work. 
We extend the signature of the data type with lambda
abstraction and application in order to express functions.
For example,
\[ \lambda x.x+x
\]
is the function that doubles its argument,
and
\[ (\lambda x.x+x)2
\]
is the function applied to argument $2$.
Adopting $\beta$-conversion as usual,
this reduces to $2+2$.
We also assume standard $\alpha$-conversion
(renaming of bound variables).
We further assume for each arity a set of function
variables.
If $f$ is a function variable of arity $k$,
we write
\[ f(t_1,\ldots,t_k)
\]
for the application of $f$ to 
arguments $t_1,\ldots,t_k$.
We write $\lambda \bar{x}.t(\bar{x})$
for the lambda abstraction
over some given, implicit number of variables $x$,
and $f(\bar{x})$ for the application
of $f$ to arguments $\bar{x}$,
where the number of arguments is
always assumed to be equal to the 
arity of $f$.

A function definition
\[ f=\lambda \bar{x}.t(\bar{x}),
\]
where $f$ is a function
variable, is expressed in the Tuplix Calculus
by the construct
\[
  \Gamma(f,\lambda \bar{x}.t(\bar{x})),
\]
and we would have, e.g.,
\[
  \Gamma(f,\lambda x.x+x)\conjc a(f(1))=
  \Gamma(f,\lambda x.x+x)\conjc a(2).
\]
To derive such identities we adopt the
axiom scheme
\begin{equation}
\tag{\axname{FD}}
  \Gamma(f,\lambda \bar{x}.t(\bar{x}))
  =
    \Gamma(f,\lambda \bar{x}.t(\bar{x}))
    \conjc\ztest{f(\bar{s})-t(\bar{s})},
\end{equation}
for any data terms $\bar{s}$.

Final step: we extend Tuplix Calculus with 
summation $\gsum_f$
over function variables $f$.
This is very similar to summation over data variables.

With these features we can define
and use functions in a `let-like' manner in specifications.
The general form
\[
  \gsum_f( \Gamma(f,\lambda \bar{x}.t(\bar{x}))
           \conjc P)
\]
may be read as `let $f$ be defined as
$\lambda \bar{x}.t(\bar{x})$ in tuplix $P$.'

For an example application we refer to~\cite{UvABAM}.
In that paper we define a budget allocation
to faculties at a university-level.
The allocation for a faculty $F$ can be given by a
faculty-independent function $f$,
which takes as input a number of parameter values
specific to $F$.
So, say that
\[
  \Gamma(f,\lambda \bar{x}.t(\bar{x}))
\]
defines $f$,
and that the allocation to $F$ is defined as
$f(\bar{x}_F)$.
The total of budget allocations is then specified by
\[
  \gsum_f(\Gamma(f,\lambda \bar{x}.t(\bar{x}))\conjc
    \Conjc_F\, (a_F(f(\bar{x}_F)))),
\]
where $a_F$ is a channel name used in the transfer of
money to $F$.

\appendix

\section{Derivations}
\label{sec:der}
Note: a zero test
$\ztest{t-s}$ may be written as
$\ztest{t=s}$.

Derivation for Example~\ref{ex:tfb}:
\begin{align*}
B &=
   \Enc_{\{a,b_1,b_2\}}
   (S\conjc Q\conjc P_1\conjc P_2)\\
  &=
   \Enc_{\{a,b_1,b_2\}}
   (&\\
   &\qquad
    a(\reward\cdot(n_1+n_2)) \conjc{}\\
   &\qquad 
     \gsum_{u} ( a(-u)\conjc c(k\cdot u)\conjc
                 (1-k)\cdot (b_1(u/2)\conjc b_2(u/2)))
     \conjc{}\\
   &\qquad \gsum_{u} (b_1(-u)\conjc d_1(u)) \conjc{}\\
   &\qquad \gsum_{u} (b_2(-u)\conjc d_2(u))
   )\displaybreak[0]\\
  &= \gsum_{u,v,w} \Enc_{\{a,b_1,b_2\}}
   (&\\
   &\qquad
    a(\reward\cdot(n_1+n_2)) \conjc{}\\
   &\qquad 
     a(-u)\conjc c(k\cdot u)\conjc
                 (1-k)\cdot (b_1(u/2)\conjc b_2(u/2))
     \conjc{}\\
   &\qquad b_1(-v)\conjc d_1(v) \conjc{}\\
   &\qquad b_2(-w)\conjc d_2(w)
   )\displaybreak[0]\\
  &=\gsum_{u,v,w} (
    \ztest{u=\reward\cdot(n_1+n_2)} \conjc{}\\
   &\qquad \ztest{v=(1-k)u/2}\conjc{}\\
   &\qquad \ztest{w=(1-k)u/2}\conjc{}\\
   &\qquad 
     c(k\cdot u)\conjc d_1(v) \conjc d_2(w))\displaybreak[0]\\
  &=
  \gsum_u ( 
   \ztest{u = 
	 \reward\cdot(n_1+n_2)}\conjc{}\\
  &\qquad
    c(k\cdot u)\conjc{}\\
  &\qquad
    (1-k)\cdot (d_1(u/2)\conjc d_2(u/2)))
\end{align*}

Derivation for Example~\ref{ex:resbuff}:
\begin{align*}
P_0 &= \Enc_{\{a_1,b_1,c_0\}}(Q_0\conjc R_0\conjc R_{1})
\displaybreak[0]\\
&=\Enc_{\{a_1,b_1,c_0\}}(\\
&\qquad
  c_0(-\pw)\conjc 
  d_0(-\inc_0)\conjc
  b_{1}(k\cdot \inc_0)\conjc
  e_0(\pw+(1-k)\cdot \inc_0) \conjc{}\\
&\qquad
  \gsum_{u,v,w,x} 
  \ztest{u+v=w+x}\conjc{}\\
&\qquad\qquad
  a_0(-u)\conjc 
  b_0(-v)\conjc
  c_0(w)\conjc
  a_1(x)\conjc{}\\
&\qquad
  \gsum_{u',v',w',x'} 
  \ztest{u'+v'=w'+x'}\conjc{}\\
&\qquad\qquad
  a_1(-u')\conjc 
  b_1(-v')\conjc
  c_1(w')\conjc
  a_2(x'))\displaybreak[0]\\
&= \gsum_{u,u',v,v',w,w',x,x'}{}\\
&\qquad  \ztest{u+v=w+x}\conjc\ztest{u'+v'=w'+x'}\conjc{}\\
&\qquad   
  d_0(-\inc_0)\conjc
  e_0(\pw+(1-k)\cdot \inc_0) \conjc{}\\
&\qquad
  a_0(-u)\conjc 
  b_0(-v)\conjc
  c_1(w')\conjc
  a_2(x')\conjc{}\\
&\qquad\Enc_H(
 c_0(-\pw)\conjc b_1(k\cdot \inc_0)\conjc{}\\
&\quad\qquad\quad
  c_0(w)\conjc
  a_1(x)\conjc 
  a_1(-u')\conjc 
  b_1(-v'))\displaybreak[0]\\
&= \gsum_{u,u',v,v',w,w',x,x'}{}\\
&\qquad  \ztest{u+v=w+x}\conjc\ztest{u'+v'=w'+x'}\conjc{}\\
&\qquad
  d_0(-\inc_0)\conjc
  e_0(\pw+(1-k)\cdot \inc_0) \conjc{}\\
&\qquad
  a_0(-u)\conjc 
  b_0(-v)\conjc
  c_1(w')\conjc
  a_2(x')\conjc{}\\
&\qquad\ztest{w=\pw}\conjc\ztest{v'=k\cdot \inc_0}\conjc
  \ztest{x=u'}\displaybreak[0]\\
&= \gsum_{u,v,w',x'}{}\\
&\qquad  
 \ztest{
   u+v=\pw+w'+x'-k\cdot \inc_0}\conjc{}\\
&\qquad
  d_0(-\inc_0)\conjc
  e_0(\pw+(1-k)\cdot \inc_0) \conjc{}\\
&\qquad
  a_0(-u)\conjc 
  b_0(-v)\conjc
  c_1(w')\conjc
  a_2(x') \displaybreak[0]\\
&= \gsum_{u,v,w,x}{}\\
&\qquad  
 \ztest{u+v=w+x+\pw-k\cdot \inc_0}
 \conjc{}\\
&\qquad
  a_0(-u)\conjc 
  b_0(-v)\conjc
  d_0(-\inc_0) \conjc{}\\
&\qquad
  a_2(x)\conjc
  c_1(w)\conjc
  e_0(\pw+(1-k)\cdot \inc_0) \displaybreak[0]\\
&= \FCO(\gsum_{u,v,w,x}{}\\
&\qquad
  a_0(-u)\conjc 
  b_0(-v)\conjc
  d_0(-\inc_0) \conjc{}\\
&\qquad
  a_2(x)\conjc
  c_1(w)\conjc
  e_0(\pw+(1-k)\cdot \inc_0))
\displaybreak[0]
\\[2ex]
%%%%%
P_1 &= 
  \Enc_{\{a_2,b_2,c_1\}}(P_0\conjc Q_1\conjc R_2)
  \displaybreak[0]\\
&= \Enc_{\{a_2,b_2,c_1\}}(\\
&\qquad \gsum_{u,v,w,x}{}\\
&\qquad\qquad  
 \ztest{u+v=w+x+\pw-k\cdot \inc_0}
 \conjc{}\\
&\qquad\qquad
  a_0(-u)\conjc 
  b_0(-v)\conjc
  d_0(-\inc_0) \conjc{}\\
&\qquad\qquad
  a_2(x)\conjc
  c_1(w)\conjc
  e_0(\pw+(1-k)\cdot \inc_0)\conjc{}\\
&\qquad
 c_1(-\pw)\conjc 
  d_1(-\inc_1)\conjc{}\\
&\qquad\qquad
  b_2(k\cdot \inc_1)\conjc
  e_1(\pw+(1-k)\cdot \inc_1) \conjc{}\\
&\qquad
 \gsum_{u,v,w,x} \\
&\qquad\qquad  \ztest{u+v=w+x}\conjc{}\\
&\qquad\qquad
  a_2(-u)\conjc 
  b_2(-v)\conjc
  c_2(w)\conjc
  a_3(x)
 )
 \displaybreak[0]\\
&=\gsum_{u,v,w,x,u',v',w',x'}\\
&\qquad
 \ztest{u'=x} \conjc
 \ztest{v'=k\cdot \inc_1} \conjc
 \ztest{w=\pw}\conjc{}\\
&\qquad  
 \ztest{u+v=w+x+\pw-k\cdot \inc_0}
 \conjc{}\\
&\qquad
  a_0(-u)\conjc 
  b_0(-v)\conjc
  d_0(-\inc_0) \conjc{}\\
&\qquad\qquad
  e_0(\pw+(1-k)\cdot \inc_0)\conjc{}\\
&\qquad
  d_1(-\inc_1)\conjc{}\\
&\qquad\qquad
  e_1(\pw+(1-k)\cdot \inc_1) \\
&\qquad
  \ztest{u'+v'=w'+x'}\conjc{}\\
&\qquad\qquad
  c_2(w')\conjc
  a_3(x') \displaybreak[0]\\
&=\gsum_{u,v,x,w',x'}\\
&\qquad
 \ztest{u+v=x+2\pw-k\cdot \inc_0}
 \conjc{}\\
&\qquad
  \ztest{x+k\cdot \inc_1=w'+x'}\conjc{}\\
&\qquad
  a_0(-u)\conjc 
  b_0(-v)\conjc
  d_0(-\inc_0) \conjc{}\\
&\qquad\qquad
  e_0(\pw+(1-k)\cdot \inc_0)\conjc{}\\
&\qquad
  d_1(-\inc_1)\conjc{}\\
&\qquad\qquad
  e_1(\pw+(1-k)\cdot \inc_1) \\
&\qquad
  c_2(w')\conjc
  a_3(x')
 \displaybreak[0]\\
&=\gsum_{u,v,w,x}\\
&\qquad
 \ztest{u+v=w+x+2\pw-k\cdot (\inc_0+\inc_1)}
 \conjc{}\\
&\qquad
  a_0(-u)\conjc 
  b_0(-v)\conjc{}\\
&\qquad  d_0(-\inc_0) \conjc
  e_0(\pw+(1-k)\cdot \inc_0)\conjc{}\\
&\qquad
  d_1(-\inc_1)\conjc
  e_1(\pw+(1-k)\cdot \inc_1) \\
&\qquad
  c_2(w)\conjc
  a_3(x)
 \displaybreak[0]\\
&=\FCO(\gsum_{u,v,w,x}\\
&\qquad
  a_0(-u)\conjc 
  b_0(-v)\conjc{}\\
&\qquad  d_0(-\inc_0) \conjc
  e_0(\pw+(1-k)\cdot \inc_0)\conjc{}\\
&\qquad
  d_1(-\inc_1)\conjc
  e_1(\pw+(1-k)\cdot \inc_1) \\
&\qquad
  c_2(w)\conjc
  a_3(x))
\end{align*}

\section{Primer on Tuplix Calculus}\label{sec:TC}
This appendix is an excerpt from~\cite{TC}.
For further reading on meadows we refer to~\cite{BT07,GBT}.
We remark that the operators
$+$ for alternative composition
and $\Enc_H$ for encapsulation
stem from the process algebra ACP~\cite{BK84}, see
also~\cite{BW,Wan}.
The summation operator $\gsum$
(binding of data variables
that generalizes alternative composition)
is also part of the specification
language $\mu$CRL~\cite{GP95},
which combines ACP
with equationally specified abstract data types.

\subsection{Cancellation Meadows}\label{sec:ZTF}
Tuplix Calculus builds on a
data type for \emph{quantities}.
This data type is required to be a
\emph{non-trivial cancellation meadow}, or,
equivalently, a \emph{zero-totalized field\/}~\cite{BT07,GBT}. 
A zero-totalized field is 
the well-known algebraic structure `field'
with a total operator for division so 
that the result of division 
by zero is zero
(and, for example,
in a 47-totalized field one has chosen 47 to represent 
the result of all divisions by zero). 

A \emph{meadow} is a commutative ring with unit equipped with
a total unary operation $(\_)^{-1}$
named inverse that satisfies the axioms
\[
  (x^{-1})^{-1} = x   \quad\text{and}\quad
  x\cdot(x \cdot x^{-1}) = x,
\]
and in which $0^{-1}=0$.
For Tuplix Calculus we also require the 
\emph{cancellation axiom}
\[
  x\neq 0 \quad \& \quad
  x\cdot y=x\cdot z\quad\Rightarrow\quad y=z
\]
to hold,
thus obtaining \emph{cancellation meadows}, which we take as
the mathematical structure for quantities, requiring further
that $0\neq 1$ to exclude (trivial) one-point models.
These axioms for cancellation meadows
characterize exactly
the equational theory of zero-totalized fields~\cite{GBT}.
The property of cancellation meadows that is
exploited in the Tuplix Calculus is
that division by zero yields zero, 
while $x\cdot x^{-1}=1$ for $x\neq 0$.

We define a
\emph{data type} (signature and axioms) for quantities
which comprises the constants 
0, 1, the binary operators $+$ and $\cdot$,
and the unary operators
$-$ and $(\_)^{-1}$.
We often write
$x-y$ instead of $x+(-y)$,
$x/y$ instead of $x\cdot y^{-1}$, and
$xy$  instead of $x\cdot y$,
and we shall omit brackets if no confusion can arise
following the usual binding conventions.
Finally, we use numerals in the common way
(2 abbreviates $1+1$, etc.).
The axiomatization consists of the cancellation axiom
\[
  x\neq 0 \quad \& \quad x\cdot y=x\cdot z 
   \quad\Rightarrow\quad y=z,
\]
the \emph{separation axiom} 
\[
  0\neq 1,
\]
and the following 10 axioms for meadows (see \cite{GBT}):
\begin{align*}
	(x+y)+z &= x + (y + z),\\
	x+y     &= y + x .\\
	x+0     &= x ,\\
	x+(-x)  &= 0 ,\\\displaybreak[0]
	(x \cdot y) \cdot  z &= x \cdot  (y \cdot  z),\\
	x \cdot  y &= y \cdot  x,\\
	1\cdot x &= x, \\\displaybreak[0]
	x \cdot  (y + z) &= x \cdot  y + x \cdot  z ,\\
	(x^{-1})^{-1} &= x ,\\
	x \cdot (x \cdot x^{-1}) &= x.
\end{align*}

The following identities are derivable from the axioms
for meadows.
\begin{align*}
	(0)^{-1}  &= 0\\
	(-x)^{-1} &= -(x^{-1})\\
	(x \cdot  y)^{-1} &= x^{-1} \cdot  y^{-1}\\
	0\cdot x  &= 0\\
	x\cdot -y &= -(x\cdot y)\\
	-(-x)     &= x
\end{align*}
Furthermore,
the cancellation axiom
and axiom $x \cdot (x \cdot x^{-1}) = x$ 
imply the \emph{general inverse law} 
\[
  x\neq 0 \quad\Rightarrow\quad x\cdot x^{-1}=1
\]
of zero-totalized fields.

\subsection{Basic Tuplix Calculus}
Core Tuplix Calculus (CTC)
is parametrized with a nonempty set \Attr\
of \emph{attributes}.
Its signature contains the
constants \et\ (the empty tuplix)
and \nt\ (the null tuplix),
and two further kinds of atomic tuplices:
\emph{entries} (attribute-value pairs) of the form
\[ a(t)
\]
with $a\in\Attr$, and $t$ a data term, 
and, for any data term $t$, the \emph{zero test}
\[ \ztest{t}
\]
($\gamma\not\in\Attr$).
Finally, CTC has one binary infix operator:
the \emph{conjunctive composition} operator \conjc.
This operator is commutative and associative.
Axioms are in Table~\ref{tab:ax}.

In CTC, a tuplix is 
a conjunctive composition of tests and entries,
with \et\ representing an empty tuplix,
and \nt\ representing an erroneous situation
which nullifies the entire composition.
Entries with the same attribute can be combined
to a single entry containing the sum of the quantities
involved.

\begin{table}[thp]
\caption{Axioms for Basic Tuplix Calculus}\label{tab:ax}
\begin{center}
\begin{minipage}{7cm}
\hrule\small
\begin{align}
\tag{\axname{T1}}\label{ax:T1} X\conjc Y &= Y\conjc X\\
\tag{\axname{T2}}\label{ax:T2} 
	(X\conjc Y)\conjc Z &= X\conjc(Y \conjc Z)\\
\tag{\axname{T3}}\label{ax:T3} X\conjc\et &= X\\
\tag{\axname{T4}}\label{ax:T4} X\conjc\nt &= \nt\\\displaybreak[0]
\tag{\axname{T5}}\label{ax:T5} a(x)\conjc a(y) &= a(x+y)\\
\tag{\axname{T6}}\label{ax:T6}\ztest x &= \ztest{x/x}\\
\tag{\axname{T7}}\label{ax:T7}\ztest 0 &= \et\\
\tag{\axname{T8}}\label{ax:T8}\ztest 1 &= \nt\\
\tag{\axname{T9}}\label{ax:T9}
	\ztest{x}\conjc\ztest{y}&= \ztest{x/x+y/y}\\
\tag{\axname{T10}}\label{ax:T10}
	\ztest{x-y}\conjc a(x)&= \ztest{x-y}\conjc a(y)
\\[2ex]
\tag{\axname{C1}}\label{ax:C1} X+Y    &= Y+X\\
\tag{\axname{C2}}\label{ax:C2}(X+Y)+Z &= X+(Y+Z)\\
\tag{\axname{C3}}\label{ax:C3} X+X    &= X\\
\tag{\axname{C4}}\label{ax:C4} X+\nt  &= X\\
\tag{\axname{C5}}\label{ax:C5}
	X\conjc (Y+Z) &= (X\conjc Y)+(X\conjc Z)\\
\tag{\axname{C6}}\label{ax:C6}
	\ztest{x}+\ztest{y} &= \ztest{xy}
\end{align}
\hrule
\end{minipage}
\end{center}
\end{table}

A zero test \ztest{t} acts as a conditional:
if the argument $t$ equals zero,
then the test is void and disappears
from conjunctive compositions.
If the argument is not equal to zero,
the test nullifies
any conjunctive composition containing it.
Observe how we exploit the property
of zero-totalized fields that $t/t$ is
always defined, and that the division $t/t$ yields 
zero if $t$ equals zero, and 1 otherwise.
Further note that an equality test $t=s$ can be
expressed as \ztest{t-s}.

A tuplix term is \emph{closed}
if it is does not contain tuplix variables and
also does not contain data variables.
A tuplix term is \emph{tuplix-closed}
if it does not contain tuplix variables
(but it may contain data variables).

The tuplix calculus is two-sorted.
On the tuplix side we have the axioms 
\ref{ax:T1}--\ref{ax:T10}
and we use the proof rules of equational logic.
On the data side, we refrain from
giving a precise proof theory.
The rule~\ref{ax:DE}
lifts valid data identities
to the tuplix calculus:
for all (open) data terms $t$ and $s$,
\begin{equation}
\label{ax:DE}
\tag{\DE}
  \DM\models t=s
  \quad\text{implies}\quad
  \ztest t =\ztest{s}, 
\end{equation}
where \DM\ (a non-trivial cancellation meadow)
is our model of the data type.
This axiom system with axioms
\ref{ax:T1}--\ref{ax:T10} plus
proof rule~\ref{ax:DE} is denoted by CTC.

The axiom system CTC is extended to
{Basic Tuplix Calculus} (BTC),
by addition of the binary operator $+$ called 
\emph{alternative composition}
or \emph{choice} to the signature,
and by adoption of axioms
\ref{ax:C1}--\ref{ax:C6}
(see Table~\ref{tab:ax}).

The following two proof rules
are derivable:
\[
  \DM\models t = s
  \quad\text{implies}\quad
  P[t/x] = P[s/x],
\]
and
\[
  P\conjc\ztest{x-t} = P[t/x]\conjc\ztest{x-t},
\]
for tuplix terms $P$ and
with substitution $P[t/x]$ defined as usual
for two-sorted equational logic
(replacement of all data variables $x$ in $P$ by $t$).

\subsection{Zero-Test Logic}
\label{sec:ZTL}
We present some observations on the use of the
zero-test operator which lead to a simple logic.

First, the empty tuplix \et\
with $\et=\ztest 0$ by axiom \ref{ax:T7}
may be read as `true', and
the null tuplix \nt\ 
with $\nt=\ztest 1$ by axiom \ref{ax:T8}
may be read as `false'.

Negation.
Define the test `not $x=0$' by
\[
  \nztest{x}\defeq \ztest{1-x/x}.
\]

Conjunctive composition of tests may be read as
logical conjunction:
\[
  \ztest{x}\conjc\ztest{y} 
   \stackrel{\eqref{ax:T9}}{=}
  \ztest{x/x+y/y}
\]
tests `$x=0$ and $y=0$'.

Alternative composition of tests may be read as
logical disjunction:
\[
  \ztest x+\ztest y
  \stackrel{\eqref{ax:C6}}{=}
  \ztest{x\cdot y}
\]
tests `$x=0$ or $y=0$'.

A \emph{formula} would then be 
a tuplix-closed (no tuplix variables)
BTC term without entries.
Any formula can be expressed as a single test 
\ztest{t} 
using axioms \ref{ax:T7}--\ref{ax:T9} and
\ref{ax:C6}, and the definition of negation.
We find that this logic has all the usual properties.
Clearly, conjunction and disjunction are
commutative, associative, and idempotent,
and it is not difficult to derive
distributivity, absorption, and
double negation elimination.
As usual, implication can be defined
in terms of negation and disjunction:
\[
  \nztest{x}+\ztest{y} = 
  \ztest{(1-x/x)\cdot y}
\]
tests `$x=0$ implies $y=0$'.

\subsection{Generalized Alternative Composition
  and Auxiliary Operators}
\label{sec:gensum}
The \emph{generalized alternative composition
(\emph{or:} summation) operator} $\gsum_x$
is a unary operator that 
\emph{binds} data variable $x$ and can be seen
as a data-parametric generalization of 
the alternative composition operator $+$.
We add this binder to the signature of BTC
and write $\FV(P)$ for the set of free data 
variables occurring in tuplix term $P$.
We write $\V(t)$ for the set of data 
variables occurring in data term $t$
(there is no variable binding within data terms).
Define substitution $P[t/x]$ as: replace
every free occurrence of data variable $x$
in tuplix term $P$ by the data term $t$,
such that no variables of $t$ become bound
in these replacements.
E.g., recall the proof rule
\[
  P\conjc\ztest{x-t} = P[t/x]\conjc\ztest{x-t}.
\]
This rule remains sound in the setting with summation,
but application of the rule may
require the renaming of bound variables in $P$,
so that the substitution can be performed.
When considering substitutions we implicitly
assume that bound variables are renamed properly.
The axiom schemes for summation are
listed in Table~\ref{tab:aux}.

\paragraph{Auxiliary Operators.}
For BTC with summation, 
we define three auxiliary operators:
scalar multiplication, clearing, and encapsulation.
Axioms are listed in Table~\ref{tab:aux}.

\begin{table}
\caption{Axiom schemes for generalization and
auxiliary operators.
Terms $P$ and $Q$ range over
tuplix terms and $t$ ranges over data terms.
}\label{tab:aux}
\begin{center}
\begin{minipage}{10cm}
\hrule\small
\begin{align}
\tag{\axname{S1}}\label{ax:S1}
	 \gsum_x P &= P
	&\text{if }x\not\in\FV(P)\\
\tag{\axname{S2}}\label{ax:S2}
	 \gsum_x P &= \gsum_y P[y/x]
	&\text{if }y\not\in\FV(P)\\
\tag{\axname{S3}}\label{ax:S3}
	 \gsum_x (P\conjc Q) &= P\conjc\gsum_x Q
	&\text{if }x\not\in\FV(P)\\
\tag{\axname{S4}}\label{ax:S4}
	 \gsum_x (P+Q) &= \gsum_xP + \gsum_xQ\\
\tag{\axname{S5}}\label{ax:S5}
	 \gsum_x\ztest{x-t} &= \et
	&\text{if }x\not\in\V(t)\\
\tag{\axname{S6}}\label{ax:S6}
	 \gsum_x\nztest{x-t} &= \et
	&\text{if }x\not\in\V(t)
\\[2ex]
\tag{\axname{Sc1}} x\cdot\et &= \et\\
\tag{\axname{Sc2}} x\cdot\nt &= \nt\\
\tag{\axname{Sc3}} x\cdot\ztest y &= \ztest y\\
\tag{\axname{Sc4}} x\cdot a(y) &= a(x\cdot y)\\
\tag{\axname{Sc5}}
  x\cdot (X\conjc Y) &= x\cdot X \conjc x\cdot Y\\
\tag{\axname{Sc6}}
  x\cdot (X+Y) &= x\cdot X + x \cdot Y\\
\tag{\axname{Sc7}}
  t\cdot\gsum_y P &= 
    \gsum_y(t\cdot P) &\text{if }y\not\in\V(t)
\\[2ex]
\tag{\axname{Cl1}}\Clr_I(\et) &= \et\\
\tag{\axname{Cl2}}\Clr_I(\nt) &= \nt\\
\tag{\axname{Cl3}}\Clr_I(\ztest x) &= \ztest x\\
\tag{\axname{Cl4}}\Clr_I(a(x)) &=
  \begin{cases}
   \et  & \text{if }a\in I\\
   a(x) & \text{otherwise}
  \end{cases}\\
\tag{\axname{Cl5}}
  \Clr_I(X\conjc Y) &=\Clr_I(X)\conjc\Clr_I(Y)\\
\tag{\axname{Cl6}}
  \Clr_I(X+Y) &= \Clr_I(X) + \Clr_I(Y)\\
\tag{\axname{Cl7}}
  \Clr_I(\gsum_xP) &=\gsum_x(\Clr_I(P))
\\[2ex]
\tag{\axname{E1}}\label{ax:E1}
  \Enc_H(\et) &= \et\\
\tag{\axname{E2}}\label{ax:E2}
  \Enc_H(\nt) &= \nt\\
\tag{\axname{E3}}\label{ax:E3}
  \Enc_H(\ztest{x}) &= \ztest{x}\\\displaybreak[0]
\tag{\axname{E4}}\label{ax:E4}
  \Enc_H(a(x)) &=
  \begin{cases}
   \ztest{x} & \text{if } a\in H\\
   a(x)      & \text{if } a\not\in H
   \end{cases}\\\displaybreak[0]
\tag{\axname{E5}}\label{ax:E5}
  \Enc_H(X\conjc\Enc_H(Y)) &=
  \Enc_H(X)\conjc\Enc_H(Y)\\
\tag{\axname{E6}}\label{ax:E6}
  \Enc_H(X+Y)&= \Enc_H(X)+\Enc_H(Y)\\
\tag{\axname{E7}}\label{ax:E7}
  \Enc_H(\gsum_xP) &=\gsum_x(\Enc_H(P))
\end{align}
\hrule
\end{minipage}
\end{center}
\end{table}

\begin{itemize}
\item 
Scalar multiplication $t\cdot P$
multiplies the quantities contained in entries
in tuplix term $P$ by $t$.
Axiom~\axname{Sc7} is an axiom scheme
with $t$ ranging over data terms and $P$ ranging over tuplix terms.

\item Clearing:
For set of attributes $I\subseteq \Attr$, the operator
$\Clr_I(X)$
renames all entries of $X$ with attribute in $I$ to \et.
It ``clears'' the attributes contained in $I$.
For a set of attributes $J\subseteq\Attr$
we further define
\[
  \Select_J(X)\defeq \Clr_{\Attr\setminus J}(X).
\label{def:Select}
\]
This function allows to focus on those entries with
attribute from $J$.

\item 
Encapsulation can be seen as `conditional clearing'.
For set of attributes $H\subseteq\Attr$,
the operator $\Enc_H(X)$ encapsulates all entries in $X$
with attribute $a\in H$.
That is, for $a\in H$, if the accumulation of quantities
in entries with attribute $a$ equals zero,
the encapsulation on $a$ is considered successful
and the $a$-entries are \emph{cleared\/} (become \et);
if the accumulation is not equal to zero, 
they become null (\nt).
This accumulation of quantities is
computed per alternative:
the encapsulation operator distributes over alternative
composition.

We further define
\[
  \Enc_{H\cup H'}(X)\defeq\Enc_H\circ\Enc_{H'}(X).
\]
\end{itemize}

\bibliographystyle{plain}

\begin{thebibliography}{9}

\bibitem{BW}
J.C.M. Baeten and W.P. Weijland.
\newblock {\em Process Algebra}.
\newblock Cambridge Tracts in Theoretical Computer Science 18,
Cambridge University Press, 1990.

\bibitem{BK84}
J.A. Bergstra and J.W. Klop.
\newblock Process algebra for synchronous communication.
\newblock {\em Information and Control} 60(1--3):109--137, 1984.

\bibitem{TFB}
J.A.\ Bergstra, S.~Nolst Trenit\'e and M.B.~van der Zwaag.
Towards a formalization of budgets.
\newblock
arXiv.org, 
%\href{http://arxiv.org/abs/0802.3617}
{arXiv:0802.3617v1}
[cs.LO], 2008.

\bibitem{UvABAM}
J.A.\ Bergstra, S.~Nolst Trenit\'e and M.B.~van der Zwaag.
UvA budget allocation model.
Report PRG0805, Section Software Engineering,
University of Amsterdam, 2008.

\bibitem{GBT}
J.A.\ Bergstra and A.~Ponse.
A generic basis theorem for cancellation meadows.
arXiv.org, 
%\href{http://arxiv.org/abs/0803.3969}
{arXiv:0803.3969v2}
[math.RA], 2008.

\bibitem{TC}
J.A.\ Bergstra, A.~Ponse and M.B.~van der Zwaag.
Tuplix Calculus.
arXiv.org, 
%\href{http://arxiv.org/abs/0712.3423}
{arXiv:0712.3423v1}
[cs.LO], 2007.

\bibitem{BT07}
J.A. Bergstra and J.V. Tucker.
\newblock The rational numbers as an abstract data type.
\emph{Journal of the ACM} 54(2), 2007.	 	

\bibitem{Wan}
W. Fokkink.
\newblock {\em Introduction to Process Algebra}.
Texts in Theoretical Computer Science,
Springer-Verlag, 2000.

\bibitem{GP95}
J.F. Groote and A. Ponse.
\newblock The syntax and semantics of $\mu${CRL}.
In: A. Ponse, C. Verhoef and S.F.M. van Vlijmen (editors),
{\em Algebra of Communicating Processes '94},
pages 26--62,
Workshops in Computing Series,
Springer-Verlag, 1995.

\end{thebibliography}

\end{document}